\documentclass[]{spie} 
\usepackage[]{graphicx}

\title{Direct comparison of the performance of CZT detectors contacted 
  with various metals}

\author{I. Jung\supit{a}, M. Groza\supit{b}, J. Perkins\supit{a}, H. Krawczynski\supit{a}, A. Burger\supit{b}
\skiplinehalf
\supit{a}Washington University in St. Louis, Department of Physics, 1 Brookings Dr., St. Louis, MO 63130\\
\supit{b}Fisk University, Department of Physics, 1000 Seventeenth Ave. North, Nashville, TN 37208
}
\authorinfo{Further author information: (Send correspondence to I.J.)\\I.J.: E-mail: jung@physics.wustl.edu, Telephone: 1 314 935 6254}

\begin{document} 
\maketitle 
  
\begin{abstract}
  Cadmium Zinc Telluride (CZT) achieves excellent spatial resolution and 
  good energy resolution over the broad energy range from several keV into 
  the MeV energy range. In this paper we present the results of a systematic 
  study of the performance of CZT detectors manufacturered by Orbotech (before 
  IMARAD) depending on surface preparation, contact materials and contact 
  deposition. The standard Orbotech detectors have the dimension of 
  2.0$\times$2.0$\times$0.5 cm. They have a pixellated In anode with  
  8$\times$8 pixels and a monolithic In cathode. Using the same CZT substrates
  several times, we have made a direct comparison of the performance of 
  different contact materials by replacing the cathode and/or the anode 
  contacts with several high-workfunction metals. We present the performance
  of the detectors and conclude with an overview over our ongoing detector 
  optimization.
\end{abstract}


\section{INTRODUCTION}
Cadmium Zinc Telluride (CZT) has emerged as the detector
material of choice for the detection of hard X-rays and soft gamma-rays
with excellent position and energy resolution and without the need for
cryogenic cooling. 
The main application of CZT detectors is the detection of photons in the 
10~keV to $\sim$1~MeV energy range. 
The high density of CZT ($\rho\,\approx$ 5.76 g/cm$^3$) and high 
average atomic number ($\simeq$50) result in high stopping power 
and in a large cross section for photoelectric interactions.

The two major vendors of CZT substrates are eV Products\cite{eV} and Orbotech\cite{Imarad}. 
These companies use different methods to grow the crystals. While eV Products uses the High-Pressure Bridgman (HPB) process  to grow their substrates, Orbotech uses the modified High-Pressure Bridgman (mHPB) process.
While HPB and mHPB CZT achieve similar performances at high energies, mHPB detectors give inferior results at low energies, which is mostly due to the higher resistivity of HPB crystals ($\sim$2$\cdot 10^{11}$ ohm cm) compared to
mHPB crystals ($\sim$$10^{10}$ ohm cm) that results in lower leakage currents.
The energy resolution  at photon energies below $\sim$200 keV is typically dominated 
by electronic noise and the readout noise. 
The energy resolution above $\sim$ 200 keV is dominated by crystal inhomogeneity 
and the dependence of the induced charge on the location where photon absorption takes place.
Thus, a detector performing well at low energies does not necessarily perform well at higher energies and vice versa. 
The application of CZT detectors in hard X-ray telescopes, requires a substantial improvement of the energy 
resolution at the low-energy end. That is the reason that lead us to undertake a detailed study of the 
influence that different contact materials might have on detector performance. 
Compared to our earlier work \cite{Kraw:04}, we describe here a more direct comparison 
between different contact materials by using the {\it same} detectors with different 
cathode and/or anode contacts. First we present an overview about deposition procedures used and our readout electronics. 
Then we will show, that the use of high workfunction materials reduces the leakage current 
significantly. At last we present and discuss first results of our ongoing study. 
In the following, we always quote Full Width Half Maximum (FWHM) energy resolutions. 
\label{sect:intro} 
\section{Detector Preparation and Setup}
We tested different contact materials using three 2 $\times$ 2 $\times$ 0.5 cm$^3$ CZT crystals from 
the company Orbotech. Two of the detectors had been contacted with a monolithic In cathode 
and 8 $\times$ 8 In pixels. Each pixel has a round diameter of 1.6 mm and a pixel pitch of 2.4 mm. 
In case of these two detectors (in the following called DI and DII) we changed only 
the contact material of the cathode while leaving the pixellated anode untouched. 
We removed the old cathode contacts by polishing with abrasive paper, fine polishing then with 0.5$\mu$m 
particle size alumina suspension and rinsing clean with pure Methanol. During the polishing process, 
the quality of the polished surface was constantly monitored with an optical
microscope. The metalization was performed with an electron beam evaporator.

The third detector (in the following called DIII) was a plain CZT crystal. In this case, we 
contacted both, the anode and the cathode. A single cathode contact was used and the 
anode side of DIII was pixellated with 2 $\times$ 2 quadratic pixels with a side length of 6.4~mm. 
The same procedure described above was used to remove the contacts. After polishing
the detectors were etched for two minutes in a 1\% bromine-methanol solution 
and subsequently rinsed in methanol. The contacts were deposited with a sputter system.
In Table \ref{tab:prep} a summary of the detector preparation of the individual 
detectors is given.\\[2ex]
\begin{table}[h]
\begin{center}       
\begin{tabular}{|l|l|l|l|} 
\hline
\rule[-1ex]{0pt}{3.5ex}  Detector &  Polished & Etched & Contacting \\
\hline
\rule[-1ex]{0pt}{3.5ex}   DI & 0.5$\mu$ alumina suspension & no & E-Beam \\
\rule[-1ex]{0pt}{3.5ex}   DII & 0.5$\mu$ alumina suspension & no & E-Beam \\
\rule[-1ex]{0pt}{3.5ex}   DIII & 0.5$\mu$ alumina suspension & 1\% Bromine 
& Sputtered \\
\hline 
\end{tabular}
\end{center}
\caption{Detector preparation of the individual Orbotech detectors used for the study.}
\label{tab:prep}
\end{table} 
We used two setups (one at Washington University and one at Fisk University) to test the detectors.
In the first setup, we use gold-plated, spring-loaded pogo-pin contacts to mount the CZT detectors
and to ensure good electrical contacts and to prevent mechanical damage to the contact pads.
We found that the leakage currents of most detectors increase dramatically once the bias voltage 
exceeds a certain limit. Usually the ``breakdown voltages'' lie in the range between -500~V and 
several thousand Volts. We biased the detectors at voltages between -500 V and -1500 V holding 
the anodes at ground. Usually, the higher the bias, the better the energy resolution, as long as 
no breakdown occurs. A hybrid electronic readout is used. The pulse shape is read out on 
four channels for the central anode pixels and the cathode. 
These pixels are AC coupled and amplified by a fast Amptek 250 amplifier followed by a second 
amplifier stage. After digitizing the amplified signals using a 500 MHz oscilloscope 
they are sent  via an ethernet connection to a PC. With this setup the electron drifttime 
can be determined with an accuracy of 10 ns. 
All results which we present in the following sections are obtained with this time resolved readout.  
Sixteen additional channels can be read out with an ASIC from eV Products to measure the 
pulse height information. The ASIC gives amplified and shaped signals which we digitize 
with a custom designed VME board. 
The FWHM noise of both readout chains lies between 5 keV and 10 keV. 
If not stated otherwise, we quote all energy resolutions after 
subtraction of square of the readout noise determined at zero detector bias. 
Energy spectra are taken at 122 keV and 662 keV.

In the second setup, the detectors are installed on a PC board with a ``zebra pad'' (a rubber sheet with
conducting wire bundles) that connects the anodes with the PC Board circuitry. 
As preamplifier and shaping amplifier a fast PGT, Model RG-11B/C and  a Tenelec TC 244 are used. 
The cathode is biased at a negative bias of -1500 V and the measured pixels are 
connected to 2N4416 Fet. Here, energy spectra are taken at 59 keV.
\section{ORBOTECH DETECTORS WITH ALTERNATIVE CONTACTS}
\label{sec:altcontacts}
Orbotech detectors with In anodes and cathodes show rather poor energy resolutions at low energies ($<200$~keV).
To reduce detector noise caused by leakage current, various authors reported
the use of blocking contacts on the cathode and/or anode side. Narita et al. (2000) \cite{Narita2000} reported good results for 4 mm thick Orbotech detectors. While for a standard Orbotech detector they determined an energy resolution of 6 keV they found for detectors with Au cathode and CdS anodes and for detectors with Au cathode and In anodes at 59 keV an energy resolutions of $\sim$3 keV. Nemirovski et al.\ (2001)\cite{Nemi:01}, 
tested a variety of cathode and anode contacts on Orbotech detectors.
The best energy resolution achieved was 5 keV at 122 keV with a detector
with blocking Au cathode and ohmic In anodes. 
Narita et al. (2002) \cite{Nari:02} has reported the results of a large number of detectors with Au cathodes and In anodes. Only 50\% of the detectors showed clear photopeaks. The typical energy resolution of these detectors ignoring the bad ones was  $\sim$6 keV at 122 keV. 

In the following we present results from contacting  n-type Orbotech CZT detectors with varying materials. 
As the energy resolution of different CZT substrates varies widely, we contacted only a few detectors with as many
different metals as possible. We used here mainly high workfunction metals to produce blocking
contacts on the cathode of the n-type CZT substrates. In Table \ref{tab:workfunction} the workfunction 
of the chosen metals are given. The workfunction of In is 4.12 eV. We have chosen Ti, Cr, Ni, Au and Pt 
with workfunctions 4.33 eV, 4.50 eV, 5.01 eV, 5.10 eV and 6.35 eV, respectively.
\begin{table}[h]
\begin{center}       
\begin{tabular}{|l|c|} 
\hline
\rule[-1ex]{0pt}{3.5ex}  Metal &  Workfunction [eV] \\
\hline
\rule[-1ex]{0pt}{3.5ex}   In &4.12\\
\rule[-1ex]{0pt}{3.5ex}  Ti &4.33\\
\rule[-1ex]{0pt}{3.5ex}  Cr & 4.50\\
\rule[-1ex]{0pt}{3.5ex}  Ni &5.01\\
\rule[-1ex]{0pt}{3.5ex}   Au &5.10\\
\rule[-1ex]{0pt}{3.5ex}   Pt &6.35\\
\hline 
\end{tabular}
\end{center}
\caption{Workfunction in eV for different materials used as 
anode- and cathode contacts.} 
\label{tab:workfunction}
\end{table} 

\begin{figure}[h]
  \begin{center}
    \begin{tabular}{c}
      \includegraphics[height=5.cm]{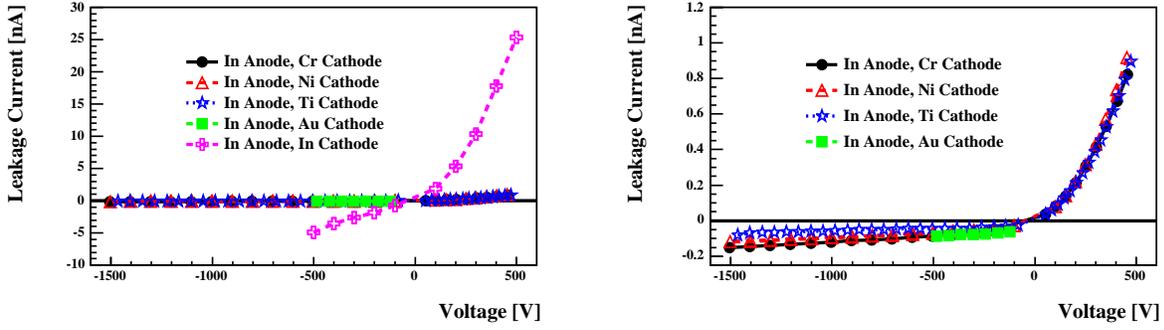}
    \end{tabular}
  \end{center}
  \caption[example]{ \label{fig:D0_IVCurve} 
    The I-V curves for Orbotech CZT detectors DI and DII with pixellated 
    In anodes with 8 $\times$ 8 pixels with a diameter of 1.6 mm and a 
    pixel pitch of 2.4 mm. The cathode materials are In (open crosses)
    Cr (filled dots), Ni (open triangles), Ti (open stars) and Au 
    (filled squares). All contact materials show a diode-like curve.     
    The graph on the right side is a zoomed part of the left side showing the finer details of the I-V curves 
    (the results for the In-In contacted detector are not shown here for clarity). 
  }
\end{figure}

\begin{figure}
  \begin{center}
    \begin{tabular}{c}
      \includegraphics[height=5.cm]{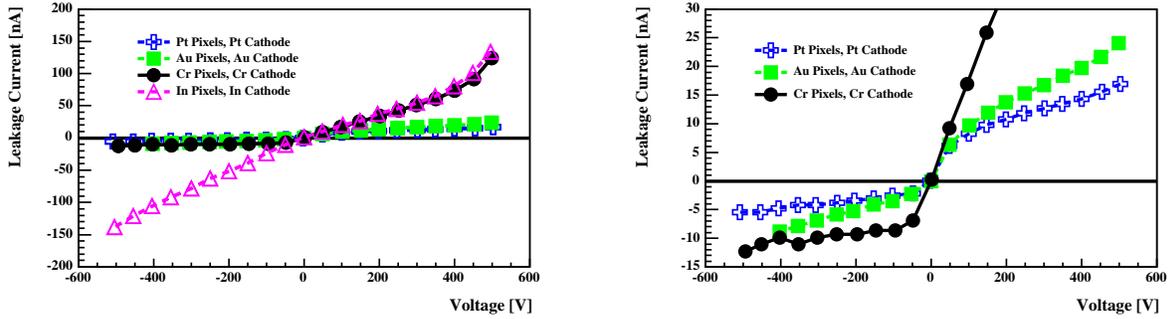}
    \end{tabular}
  \end{center}
  \caption[example]{ \label{fig:Burger_IVCurve} 
    The I-V curves for Orbotech CZT detector DIII (2 $\times$ 2 circular pixels, 
    each with a radius of 6.4~mm). 
    The anode/cathode materials are Cr/Cr (filled dots), In/In (open triangles), Pt/Pt (open crosses) 
    and Au/Au filled squares). In/In contacts show a ohmic behavior, all other contact materials show a diode-like curve. 
    The graph on the right side is a zoomed part of the left side showing the finer details of the I-V curves 
    (the results for In/In contacts are not shown here for clarity). 
}
          
\end{figure}

The I-V curves for the detectors DI and DII  are shown in  Figure \ref{fig:D0_IVCurve}. 
Both detectors have  8 $\times$ 8 In pixels and
Cr, Ni, Ti and Au cathodes. The voltage was applied to the cathode. Pixels which are not measured are held at ground potential. All curves show a diode-like behavior with leakage currents less than 1 nA. Compared to In this is smaller by a factor of 40 to 90. At a bias voltage of -400 V we measured the following leakage currents:  In -3.62 nA, Cr -0.08 nA, Ni -0.06 nA, Ti -0.04 nA and Au -0.09 nA.
In figure \ref{fig:Burger_IVCurve}  the I-V curves of the detector DIII are shown. 
The pixel-cathode currents scale approximately with the area per pixel. 
The latter was by a factor of $\sim$22 larger for detector DIII than for the detectors DI and DII.
The detector DIII with In contacts exhibited ohmic behavior, Pt, Cr and Au show diode-like behavior 
with between 50 and 100 times lower leakage currents than for In.

\begin{table}[h]
\begin{center}       
\begin{tabular}{|l|l|l|l|l|l|} 
\hline
\rule[-1ex]{0pt}{3.5ex} Detector &Anode &  Cathode & FWHM (662 keV)
& FWHM (122 keV) & Bias Voltage  \\
\hline
\rule[-1ex]{0pt}{3.5ex}DI & In & In &19.88 keV & &-1000 V\\
\rule[-1ex]{0pt}{3.5ex}DI & In & In &21.65 keV & &-1000 V\\
\rule[-1ex]{0pt}{3.5ex}DI & In & Ti &20.84 keV & 12.19 keV &-1000 V\\
\rule[-1ex]{0pt}{3.5ex}DI & In & Ti &20.77 keV & 10.32 keV &-1000 V\\
\rule[-1ex]{0pt}{3.5ex}DI & In & Ti &20.68 keV & 15.82 keV &-1000 V\\
\rule[-1ex]{0pt}{3.5ex}DI & In & Cr & 21.82 keV &14.69 keV &-500 V\\
\rule[-1ex]{0pt}{3.5ex}DI & In & Cr & 30.62 keV &13.08 keV  &-500 V\\
\rule[-1ex]{0pt}{3.5ex}DI & In & Cr & 18.01 keV &17.61 keV  &-500 V\\
\rule[-1ex]{0pt}{3.5ex}DI & In & Cr & &11.03 keV  &-1000 V\\
\rule[-1ex]{0pt}{3.5ex}DI & In & Cr & &13.59 keV  &-1000 V\\
\rule[-1ex]{0pt}{3.5ex}DI & In & Cr & &11.10 keV  &-1000 V\\
\rule[-1ex]{0pt}{3.5ex}DI & In & Ni &20.19 keV & 14.30 keV &-500 V\\
\rule[-1ex]{0pt}{3.5ex}DI & In & Ni &35.06 keV & 12.07 keV &-500 V\\
\rule[-1ex]{0pt}{3.5ex}DI & In & Ni &25.48 keV & 14.82 keV &-500V\\
\rule[-1ex]{0pt}{3.5ex}DI & In & Pt & \multicolumn{2}{|c|}{no spectra} &-500 V\\
\rule[-1ex]{0pt}{3.5ex}DI & In & Pt & \multicolumn{2}{|c|}{no spectra} &-500 V\\
\rule[-1ex]{0pt}{3.5ex}DI & In & Pt & \multicolumn{2}{|c|}{no spectra} &-500 V\\
\hline
\rule[-1ex]{0pt}{3.5ex}DII & In & In &19.73 keV & &-500 V\\
\rule[-1ex]{0pt}{3.5ex}DII & In & In &19.58 keV & &-500 V\\
\rule[-1ex]{0pt}{3.5ex}DII & In & In &28.62 keV & &-500 V\\
\rule[-1ex]{0pt}{3.5ex}DII & In & Au &17.70 keV  & &-500 V \\
\rule[-1ex]{0pt}{3.5ex}DII & In & Au &20.08 keV  & &-500 V\\
\rule[-1ex]{0pt}{3.5ex}DII & In & Au &17.71 keV  & &-500 V\\
\rule[-1ex]{0pt}{3.5ex}DII & In & Pt & \multicolumn{2}{|c|}{no spectra} &-500 V\\
\rule[-1ex]{0pt}{3.5ex}DII & In & Pt & \multicolumn{2}{|c|}{no spectra} &-500 V\\
\rule[-1ex]{0pt}{3.5ex}DII & In & Pt & \multicolumn{2}{|c|}{no spectra} &-500 V\\\hline
\end{tabular}
\end{center}
\caption{FWHM energy resolution of Orbotech detectors contacted with different metals. The Pt-In detectors did not show a clear photopeak at 122 keV and for DI and DII at 662 keV.} 
\label{tab:FWHMDIAndDII}
\end{table}

\begin{table}[h]
\begin{center}       
\begin{tabular}{|l|l|l|l|l|} 
\hline
\rule[-1ex]{0pt}{3.5ex} Detector &Anode &  Cathode & FWHM (59 keV) & Bias Voltage  \\
\hline
\rule[-1ex]{0pt}{3.5ex}DIII & Au & Au & 6.43 keV & -1500 V\\
\rule[-1ex]{0pt}{3.5ex}DIII & Cr & Cr & 15.23 keV & -1500 V\\
\rule[-1ex]{0pt}{3.5ex}DIII & Pt & Pt & 7.97 keV & -1500 V\\
\rule[-1ex]{0pt}{3.5ex}DIII & In & In & 9.85 keV & -1500 V\\
\hline
\end{tabular}
\end{center}
\caption{FWHM energy resolution of Orbotech detectors contacted with different metals.} 
\label{tab:FWHMDIII}
\end{table}

\subsection{Energy Spectra}
\begin{figure}
  \begin{center}
    \begin{tabular}{c}
      \includegraphics[height=13cm]{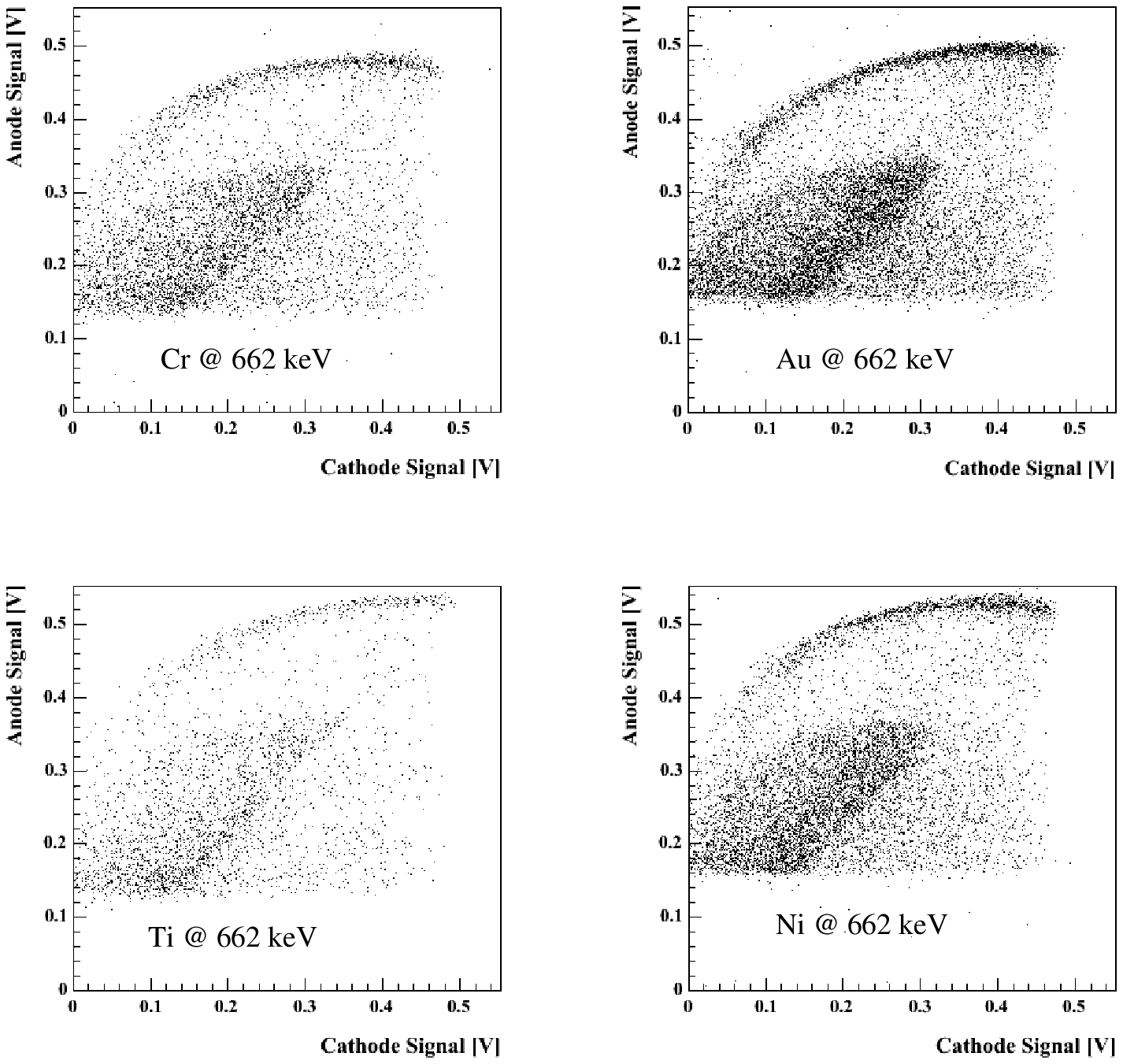}
    \end{tabular}
  \end{center}
  \caption[example]{ \label{fig:AnodeVSCathode} 
    Correlation of anode amplitude with the cathode amplitude for 662 keV for
four different cathode materials, Cr, Ni, Au and Ti. The anode material was In.}
\end{figure}

CZT is a single charge carrier: as consequence of the poor hole mobility and fast hole trapping in CZT, the induced signals are mainly produced by electrons. If the pixel size is small enough most of the signals are induced near the anode (small pixel effect \cite{He:01}).  On their way to the anodes, the electrons are trapped and the anode signal depends on the depth 
of interaction (DOI).
The cathode signal is linear with DOI providing the possibility to calculate  the DOI by using the ratio of cathode to anode signal and apply DOI corrections. The electron lifetime is $\tau_{e} = 3\mu s$, the hole lifetime $\tau_{h} = 1\mu s$, the electron and hole mobility is $\mu_{e}  = 1000$ cm$\cdot$cm/V/s and $\mu_{h}  = 65$ cm$\cdot$cm/V/s  respectively.
The dependence of the induced signals on the DOI is negligible at energies below $\sim$100 keV. Here all photon interaction takes place close to the cathode. With higher energies the importance of a DOI correction increases. We are using the cathode anode dependency 
to correct for this. In Figure \ref{fig:AnodeVSCathode} the anode signal for one central pixel is shown as a function of cathode signal for In anodes and Cr, Au, Ti and Ni cathodes. One can clearly see
the dependency.

\begin{figure}
  \begin{center}
    \begin{tabular}{c}
        \includegraphics[height=13cm]{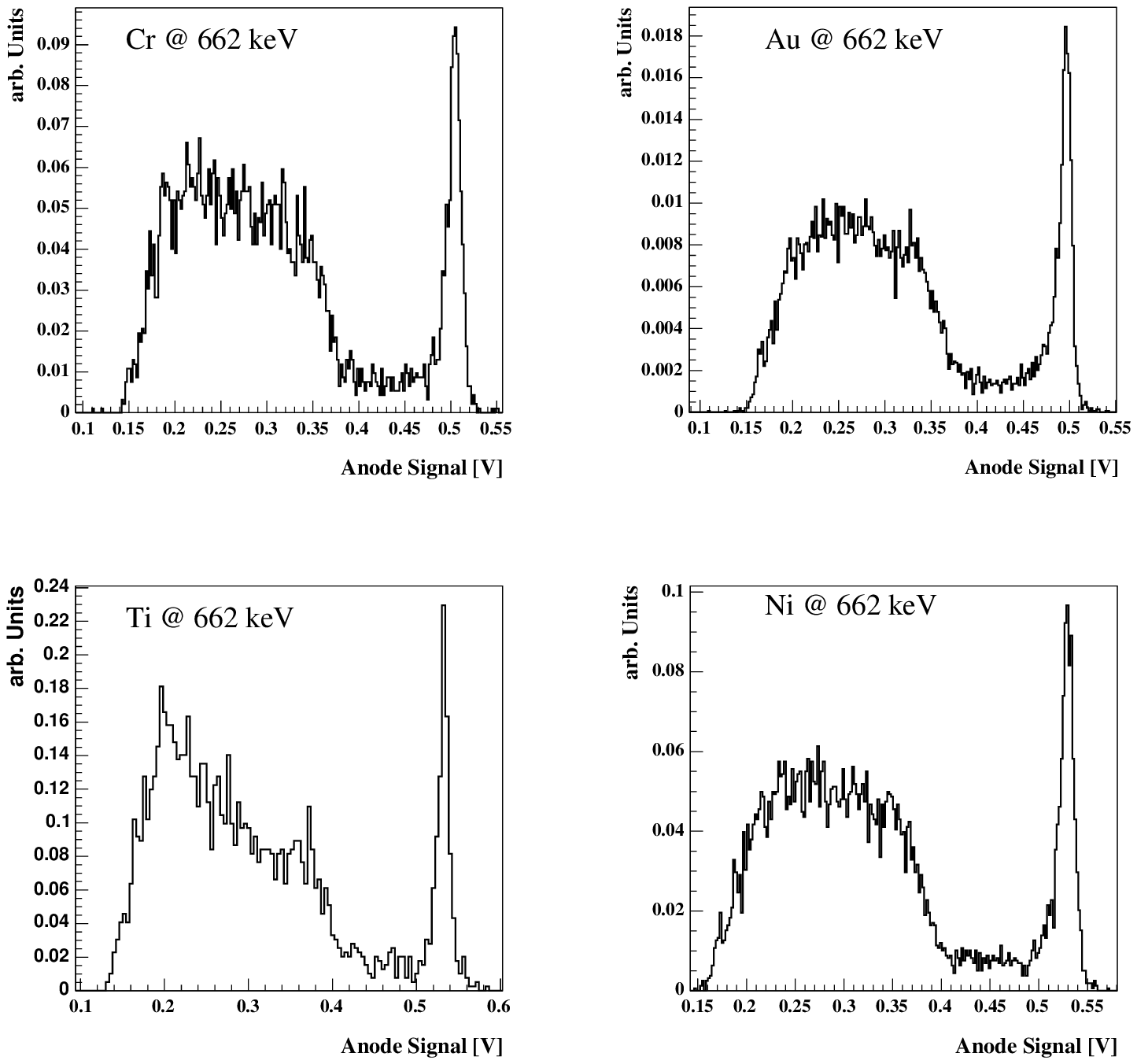}
    \end{tabular}
  \end{center}
  \caption[example]{ \label{fig:Spectra662} 
662 keV energy spectra of Orbotech detectors contacted with pixellated In
anode with 8 $\times$ 8 In pixels, each with a diameter of 1.6 mm and 
a pixel pitch of 2.4 mm and Cr, Au, Ti and Ni cathode. After 
DOI correction the FWHM energy resolutions are 18.01 keV (Cr),  17.70 keV (Au),
20.68 keV (Ti) and 20.29 keV (Ni). 
  }
\end{figure}

\begin{figure}
  \begin{center}
    \begin{tabular}{c}
      \includegraphics[height=13cm]{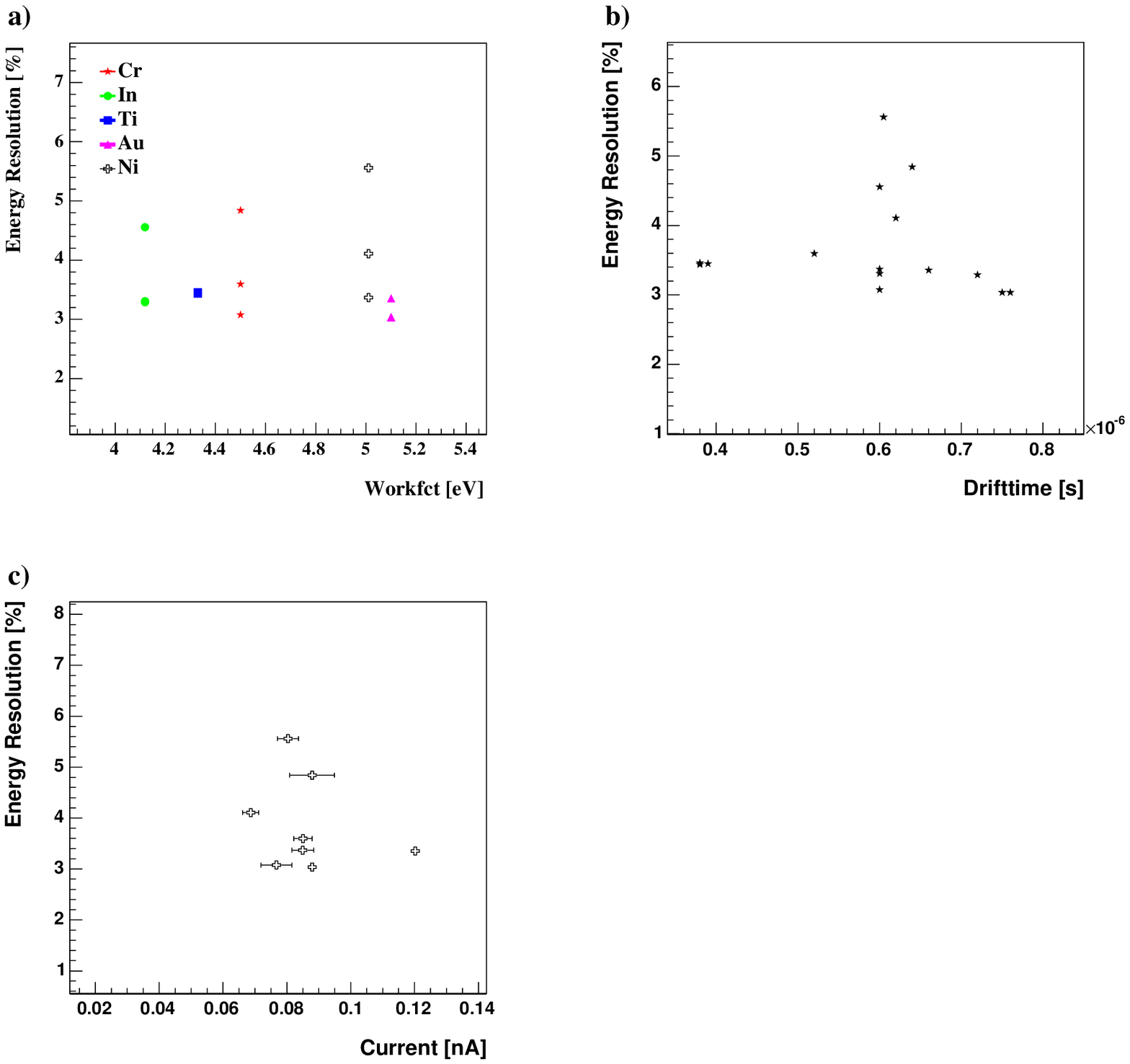}
    \end{tabular}
  \end{center}
  \caption[example]{ \label{fig:correlations} 
    Panel a) shows the 662 keV energy resolutions as a function of the workfunctions of the cathode materials; panel b) the energy resolution as a function of maximal electron drifttime and in panel c) the  energy resolution as a function
of leakage current for the detectors DI and DII.
}
\end{figure}

The energy resolutions after DOI correction are given in Table \ref{tab:FWHMDIAndDII} for detectors DI and DII. 
A set of 662 keV energy spectra are shown in Figure \ref{fig:Spectra662}. 
For detector DI, we see that the In, Ti, and Cr cathodes all give very comparable energy resolutions, about 
20 keV at 662 keV and about 12 keV at 122 keV. Ni gives substantially poorer results at 662 keV and the
Pt contacted detector did show signals, but did not show a photopeak.
For detector DII, we see that In and Au cathodes give similar results, and that Pt did not work again.
We studied the dependence of the 662 keV energy resolutions on the workfunctions of the contact materials, 
the leakage currents and the maximum observed drifttimes. The results are shown in Figure \ref{fig:correlations}. 
The study does not show any systematic effects. 
For the detector DIII, the results are shown in Table \ref{tab:FWHMDIII} at 59 keV.
Here, the Au contacts worked best, Pt and In gave reasonable results, 
and Cr showed the worst results. It can be seen that detector DIII performs considerably 
better at 59 keV (energy resolutions about 6 keV, 10\%) than the detectors 
DI and DII at 122 keV (energy resolutions of 20 keV, 16\%).
\section{SUMMARY AND OUTLOOK}
In this report, we presented the results of two ongoing studies carried through in parallel.
In the first study, we tested two detectors with standard In anode contacts and 
different cathode contacts. Although all the cathode contacts reduced the leakage current by more than 
one order of magnitude, we found that we get very similar energy resolution for In, Cr, Ti and 
Au cathodes. The two metals Ni and Pt showed markedly poorer performance. 
In the second study, we replaced both, the anode and the cathode contacts of another detector. 
In this study, Au gave the best results, followed in order of performance by Pt, In, and Cr.
In absolute terms, the detectors of the second study performed better than the detectors 
of the first study.
Our next steps in these two sudies will be to compare the detector performance with and 
without Br-etching for all three detectors DI, DII and DIII and to swap also the anode 
contacts of detectors DI and DII. The results should be able to decide if the differences 
in the performance of the detectors originate from the differences in detector fabrication 
or are simply differences in the inherent quality of the three CZT substrates.\\[2ex]
For several anode/cathode contact material combinations, we get very good results, and we have started
to test detectors with alternative contact geometries. 
The main thrust is to maximize the active volume of the detectors by using steering grids.
Direct measurements of the active volume of the present Orbotech detectors gives a relatively
low value of only 60\%. There are two ways to improve on the situation: (i) use pixels with 
larger anode pads; (ii) use steering grids to push electron clouds away from the area 
between pixels. We have started to use photolithography to fabricate detectors with
pixels separated by only $\sim$100 $\mu$, and to fabricate detectors with steering grids.
\acknowledgments  
We thank Uri El Hanany from Orbotech for several free CZT detectors. 
We acknowledge L.~Sobotka, D.~Leopold, and J.~Buckley for helpful discussions. 
Thanks to electrical engineer P.~Dowkontt, and electrical technician 
G.~Simburger for their support. HK is grateful to J.~Matteson for very 
detailed discussion of many key-characteristics of CZT detectors.
This work is supported by NASA under contracts NNG04WC176 and 
NNG04GD70G, and the NSF/HRD grant no.\ 0420516 (CREST).
\bibliography{report}   
\bibliographystyle{spiebib}   

\end{document}